\documentclass[preprint,11pt,floats,epsf,superscriptaddress,nofootinbib]{revtex4-1}
\usepackage{graphicx}
\usepackage{dcolumn} 
\usepackage{bm}
\usepackage{bbm}
\usepackage{amssymb}
\usepackage{amsmath}
\usepackage{epsfig}    
\usepackage{color}
\usepackage{xcolor}
\usepackage{slashed}
\usepackage{multirow}
\usepackage[normalem]{ulem}
\usepackage{physics}
\usepackage[colorlinks=true,
            linkcolor=red,
            filecolor=magenta,
            citecolor=blue,
            urlcolor=magenta
            ]{hyperref} 
\let\oldfootnote\footnote
\renewcommand{\footnote}[1]{%
    \begingroup%
    \linespread{1}
    \oldfootnote{#1}%
    \endgroup%
}

\newcommand{\gev}{\,\mathrm{GeV}}
\newcommand{\tev}{\,\mathrm{TeV}}
\newcommand{\citere}[1]{Ref.~\cite{#1}}
\newcommand{\citeres}[1]{Refs.~\cite{#1}}
\newcommand{\refse}[1]{Sec.~\ref{#1}}

\newcommand{\muCMS}{0.33}
\newcommand{\dmuCMSpl}{0.19}
\newcommand{\dmuCMSmi}{0.12}
\newcommand{\sigATLAS}{{1.7}}
\newcommand{\muATLAS}{{0.18}}
\newcommand{\dmuATLASpl}{{0.10}}
\newcommand{\dmuATLASmi}{{0.10}}
\newcommand{\muACgaga}{\mu_{\gamma\gamma}^{{\rm ATLAS+CMS}}}
\newcommand{\muAgaga}{\mu_{\gamma\gamma}^{{\rm ATLAS}}}
\newcommand{\muCgaga}{\mu_{\gamma\gamma}^{{\rm CMS}}}
\newcommand{\muAC}{{0.24}}
\newcommand{\dmuACp}{{0.09}}
\newcommand{\dmuACm}{{0.08}}
\newcommand{\sigAC}{{3.1}}

\newcommand{\massATLAS}{{95.4}}
\newcommand{\massAC}{{95.4}}

\allowdisplaybreaks

\begin{document}

\count\footins = 1000

\preprint{DESY-23-221, IFT--UAM/CSIC-23-066}

\title{A 95 GeV Higgs Boson in the Georgi-Machacek Model}

\author{Ting-Kuo Chen}
\email{tchen463@wisc.edu}
\affiliation{Department of Physics, University of Wisconsin-Madison, Madison, WI 53706, USA}

\author{Cheng-Wei Chiang}
\email{chengwei@phys.ntu.edu.tw}
\affiliation{Department of Physics and Center for Theoretical Physics, National Taiwan University, Taipei, Taiwan 10617, ROC}
\affiliation{Physics Division, National Center for Theoretical Sciences, Taipei, Taiwan 10617, ROC}

\author{Sven Heinemeyer}
\email{sven.heinemeyer@cern.ch}
\affiliation{Instituto de Física Teórica UAM-CSIC, Cantoblanco, 28049, Madrid, Spain}

\author{Georg Weiglein}
\email{georg.weiglein@desy.de}
\affiliation{Deutsches Elektronen-Synchrotron DESY, Notkestr.~85, 22607 Hamburg, Germany}
\affiliation{II.~Institut für Theoretische Physik, Universität Hamburg, Luruper Chaussee 149, 22607 Hamburg, Germany}

\begin{abstract}
CMS and ATLAS have reported small excesses in the search for low-mass Higgs bosons in the di-photon decay 
channel at exactly the same mass, $95.4 \gev$. These searches rely on improved analysis techniques, enhancing in particular the discrimination against
the $Z \to e^+e^-$ background. 
In models beyond the Standard Model (SM) that extend the Higgs sector with triplets, doubly-charged 
Higgs bosons are predicted which can contribute substantially to the di-photon decay rate of a light Higgs boson.
The Georgi-Machacek (GM) Model is of particular interest in this context, since despite containing Higgs triplets 
it preserves the electroweak $\rho$-parameter to be~1 at the tree level. 
We show that within the GM model, a Higgs boson with a mass of $\sim 95 \gev$ with a di-photon decay 
rate as observed by CMS and ATLAS 
can be well described. We discuss  the di-photon excess in conjunction with an excess in the
$b \bar b$ final state observed at LEP and an excess observed by CMS in the di-tau
final state, which have been found at comparable masses with local significances of about
$2\sigma$ and $3\sigma$, respectively. 
The presence of a Higgs boson at about $95 \gev$ within the GM model would imply good prospects of the searches for additional light Higgs bosons. In particular, the observed excess 
in the di-photon channel would be expected to be correlated in the GM model
with a light doubly-charged Higgs boson in the mass range between $100 \gev$ and $200 \gev$, which 
motivates dedicated searches
in upcoming LHC Runs. 
\end{abstract}

\maketitle

\newpage
\section{Introduction}\label{sec:intro}

In the year 2012, the ATLAS and CMS collaborations discovered a new scalar particle~\cite{ATLAS:2012yve,CMS:2012qbp}.
Within the current theoretical and experimental uncertainties, the properties  of the new particle are
consistent with the predictions for the Higgs boson of the Standard Model (SM) with a mass
of $\sim 125\gev$~\cite{CMS:2022dwd,ATLAS:2022vkf}. However, they are also compatible with many scenarios 
of physics beyond the SM (BSM).  
While the minimal scalar sector of the SM contains only one physical Higgs particle,
BSM scenarios often give rise to an extended Higgs sector containing additional scalar particles.
Accordingly, one of the primary objectives of the current and future LHC runs is the search for additional Higgs bosons,
which is of crucial importance for exploring the underlying physics of electroweak symmetry breaking. These additional 
Higgs bosons can have a mass above, but also below $125 \gev$. 

Searches for Higgs bosons below $125\gev$ have been performed at the LEP~\cite{OPAL:2002ifx,LEPWorkingGroupforHiggsbosonsearches:2003ing,ALEPH:2006tnd},
the Tevatron~\cite{CDF:2012wzv} and the 
LHC~\cite{CMS-PAS-HIG-14-037,CMS:2018cyk,CMS:2018rmh,ATLAS-CONF-2018-025,CMS:2022goy,ATLAS:2022abz,CMS-PAS-HIG-20-002}. Searches for di-photon resonances at the LHC are particularly intriguing and promising in this context, which is also apparent from the fact that this decay mode,
due to its comparably clean final state, constitutes one of the two discovery channels of the Higgs boson at $125\gev$.
CMS had performed searches for scalar di-photon resonances at~$8\tev$ and $13\tev$.
Results based on the $8\tev$ data and the first year of Run~2 data at $13\tev$,
corresponding to an integrated luminosity of $19.7\,\mathrm{fb}^{-1}$ and $35.9\,\mathrm{fb}^{-1}$,
respectively, showed a local excess of $2.8\,\sigma$ at $95.3 \gev$~\cite{CMS-PAS-HIG-14-037,CMS:2018cyk}.
This excess, being present in both the $8\tev$ and the $13\tev$ datasets, 
soon received widespread attention, 
see e.g.~\citeres{Cao:2016uwt,
Fox:2017uwr,
Richard:2017kot,
Haisch:2017gql,
Biekotter:2017xmf,
Domingo:2018uim,
Liu:2018xsw,
Cline:2019okt,
Biekotter:2019kde,
Aguilar-Saavedra:2020wrj}.

Subsequently, CMS published the result based on their full Run~2 dataset
employing substantially refined analysis techniques. 
The 
observed excess for a mass of $95.4 \gev$,
expressed in terms of a signal strength,
is given by~\cite{CMS-PAS-HIG-20-002}
\begin{equation}
\muCgaga =\frac{\sigma^{\rm exp} \left( gg \to \phi \to \gamma\gamma \right)}
         {\sigma^{\rm SM}\left( gg \to H \to \gamma\gamma \right)}
     = \muCMS^{+\dmuCMSpl}_{-\dmuCMSmi} \ .
\label{muCMS}
\end{equation}
Here $\sigma^{\rm SM}$ denotes the cross section for a hypothetical SM
Higgs boson at the same mass. 
Analyses using the result based on the full Run~2 data
can  be found, e.g., in \citeres{Biekotter:2023jld,Azevedo:2023zkg} 
(see also \citere{Biekotter:2023qbe} for a review). 

More recently, ATLAS presented the result based on their full Run~2 dataset~\cite{LHC-Seminar,ATLAS-CONF-2023-035}
(in the following, we refer to their ``model-dependent''
analysis, which has a higher discriminating power).
The new analysis has a substantially improved sensitivity with respect to their analysis based on 
the previously reported result utilizing $80\,\mathrm{fb}^{-1}$~\cite{ATLAS-CONF-2018-025}.
ATLAS finds an excess with a local significance of $\sigATLAS\,\sigma$ at 
precisely the same mass value as the one that was previously reported by CMS, i.e.\, at
$\massATLAS \gev$. This ``di-photon excess'' corresponds to a signal strength of
\begin{equation}
\muAgaga =\frac{\sigma^{\rm exp} \left( gg \to \phi \to \gamma\gamma \right)}
         {\sigma^{\rm SM}\left( gg \to H \to \gamma\gamma \right)}
     = \muATLAS^{+\dmuATLASpl}_{-\dmuATLASmi} \ ,
\label{muATLAS}
\end{equation}
see \citere{Biekotter:2023oen} for details.
Neglecting possible correlations, a combined signal strength of
\begin{equation}
    \mu^\mathrm{exp}_{\gamma\gamma} = \muACgaga = \muAC^{+\dmuACp}_{-\dmuACm}\,
    \label{muAC}
\end{equation}
was obtained~\cite{Biekotter:2023oen}, corresponding to an excess of $\sigAC\,\sigma$ at
\begin{equation}
  m_{\phi} \equiv m_{\phi}^{\rm ATLAS+CMS} = \massAC \gev\,.
\label{massAC}
\end{equation}

The excess in the $\gamma\gamma$ channel is 
also of interest in view of the fact that
LEP reported a local $2.3\,\sigma$ excess in the~$e^+e^-\to Z\phi(\phi\to b\bar{b})$
searches\,\cite{LEPWorkingGroupforHiggsbosonsearches:2003ing}, which is consistent with a Higgs boson with a mass of
$\massAC \gev$ and a signal strength of~\cite{Cao:2016uwt,Azatov:2012bz}
\begin{align}
\mu_{bb}^{\rm exp} = 0.117 \pm 0.057~.
\label{mubb}
\end{align}
In addition to the di-photon excess,
CMS also observed another excess compatible with a mass of $\massAC \gev$ in
the search for $pp \to \phi \to \tau^+\tau^-$~\cite{CMS:2022goy}.
While this excess is most pronounced at a mass of $100\gev$
with a local significance of $3.1\,\sigma$,
it is also well compatible with a mass
of $\massAC \gev$, with a local significance of $2.6\,\sigma$.
At $95 \gev$, the signal strength is determined to be
\begin{align}
\mu^{\rm exp}_{\tau\tau} = 1.2 \pm 0.5~.
\label{mutautau]}
\end{align}
ATLAS has not yet published a search in
the di-tau final state covering the mass range around 95~GeV.
Concerning the CMS excess in the di-tau channel it should be noted
that such a large signal strength is in some tension (depending on the model realization) with experimental bounds from
recent searches performed by CMS for the production of a Higgs boson
in association with a top-quark pair or in association with a $Z$~boson, with subsequent
decay into tau pairs~\cite{CMS-PAS-EXO-21-018}, as well as with the searches performed at LEP for the process $e^+e^-\to Z\phi(\phi\to\tau^+\tau^-)$~\cite{LEPWorkingGroupforHiggsbosonsearches:2003ing}. 

Given that all the excesses discussed above
occur at a similar mass, the interesting 
question arises 
whether they could all
be caused by the  production of a single new particle -- 
which, if confirmed by future experiments, would be a first 
unambiguous
sign of BSM physics in the Higgs sector.
This triggered activities in the literature regarding possible 
model interpretations that could account 
for the various excesses, 
see, besides the studies mentioned above, e.g.,~\citeres{
Hollik:2018yek,
Choi:2019yrv,
Biekotter:2019gtq,
Cao:2019ofo,
Biekotter:2021ovi,
Biekotter:2021qbc,
Heinemeyer:2021msz,
Biekotter:2022jyr,
Biekotter:2022abc,
Iguro:2022dok,
Kundu:2022bpy,Aguilar-Saavedra:2023tql,LeYaouanc:2023zvi,Cao:2023gkc}.

Many model interpretations discussed in the literature employed extensions of the Two Higgs Doublet Model (2HDM).
The main reason was to allow for a suppression of the $\phi b\bar b$ coupling, enhancing in this 
way the BR($\phi \to \gamma\gamma$), 
in order to provide an adequate description of
the CMS and ATLAS excesses in this channel.
However, there is a second possibility to enhance the di-photon decay rate. Additional charged 
particles in the loop-mediated decay $H_1 \to  \gamma\gamma$ can yield a positive contribution 
to the decay rate and in this way result in a sufficiently strong rate. 
However, as was found in \citere{Biekotter:2019kde}, a second Higgs doublet, providing an additional singly-charged scalar, is not sufficient to yield a relevant effect on BR($\phi \to \gamma\gamma$). 

The above-mentioned situation could be different in models with multiply-charged scalars.  For example, doubly-charged Higgs bosons exist in models with Higgs triplet fields and can potentially enhance BR($\phi \to \gamma\gamma$).  
In view of the constraint arising from
the electroweak $\rho$-parameter, one is led to the Georgi-Machacek (GM) model~\cite{Georgi:1985nv,Chanowitz:1985ug}, which has $\rho^{\rm tree} = 1$ by construction. 
This is realized by imposing a global $SO(4)\cong SU(2)_L\times SU(2)_R\cong SU(2)_V\times SU(2)_A$ symmetry to the extended Higgs potential. After electroweak symmetry breaking, the custodial $SU(2)_V$ symmetry is preserved at tree level, and thus contributions to $\rho$ will only 
arise at loop levels. The model also has the capability of providing Majorana mass to neutrinos through the lepton number-violating couplings between the complex Higgs triplet and lepton fields.
Moreover, it predicts the existence of several Higgs multiplets, whose mass eigenstates form one quintet ($H_5$), one triplet ($H_3$), and two singlets ($H_1$ and $h$) under the custodial symmetry. 

In this paper, we analyze the GM model with respect to the various excesses found around 
$\sim 95 \gev$.
Previous studies focused on extensions of the GM model~\cite{Kundu:2022bpy,LeYaouanc:2023zvi}. Most recently, 
while finalizing this manuscript, \citere{Ahriche:2023wkj} appeared, investigating the excesses at $\sim 95 \gev$
in the GM model. In that work, contrary to our analysis (see the detailed discussion below), 
several Higgs bosons of the GM model are assumed to
have masses around $\sim 95 \gev$, resulting in overlapping signals and leading potentially to somewhat 
larger rates in the di-tau channel. On the other hand, as will be explained below, our treatment of the 
various Higgs-boson exclusion bounds and LHC rate measurements includes all relevant data, in particular 
also from Simplified Template Cross Sections (STXS) measurements. The incorporation of the latest results 
regarding the measured properties 
of the detected Higgs boson at $\sim 125\gev$ as well as of a comprehensive set of the existing limits from additional Higgs searches 
at the LHC and previous colliders can be particularly relevant for the case of several light Higgs bosons. 
Furthermore, we include here an analysis of future explorations of the GM interpretation of the $\sim 95 \gev$ excesses.

As will be illustrated below, we identify $h$ with the discovered Higgs boson at about 125~GeV, and $H_1$ with the putative 
Higgs boson state at 95~GeV. Because of mixing among the Higgs doublet and triplet fields, the GM model presents a richer 
Higgs phenomenology through the couplings of the SM fermions and gauge bosons with the various Higgs boson states.
We will furthermore demonstrate that, as a consequence of the fact that the scale of all the exotic scalar masses 
(i.e., the masses of all the additional Higgs bosons besides the one at about $125\gev$)
are determined by the triplet mass parameter $m_2^2$ (see Eq.~\eqref{eq:potential} below)
and taking into account
the theoretical and currently available experimental constraints, the existence of the state $H_1$ at about 95~GeV 
implies that all the exotic Higgs bosons have masses at the electroweak scale.
We will show that this gives rise to very good prospects for probing this scenario with
future results from the LHC experiments\footnote{A $\mathbb{Z}_2$-symmetric version of the GM model has been studied in Ref.~\cite{deLima:2022yvn}, which may be compatible with $H_1$ around 95~GeV as well as $H_3$ and $H_5$ at the electroweak scale.}.

The paper is organized as follows.
After a brief review of the model in \refse{sec:GM-model} we list all relevant 
theoretical and experimental constraints in \refse{sec:numerical} and 
describe our corresponding analysis flow. 
The results are presented in \refse{sec:95-results}, and the future expectations of how such a scenario can be further 
analyzed at the HL-LHC or a future $e^+e^-$ collider are discussed in \refse{sec:future}. Our conclusions are given
in \refse{sec:conclusions}. 

\section{The Georgi-Machacek Model}\label{sec:GM-model}

In this section, we give a brief overview of the GM model and introduce our notation. The scalar sector of the GM model comprises one isospin doublet field ($\phi$) with hypercharge $Y=1/2$, one isospin triplet field ($\chi$) with $Y=1$, and one isospin triplet field ($\xi$) with $Y=0$.  Under the global $SU(2)_L\times SU(2)_R$ symmetry that is realized in the GM model, they can be arranged into the following covariant forms
\begin{equation}
	\Phi = \begin{pmatrix}
		(\phi^0)^* & \phi^+ \\
		-\phi^- & \phi^0
	\end{pmatrix} ,~ \Delta = \begin{pmatrix}
		(\chi^0)^* & \xi^+ & \chi^{++} \\
		-\chi^- & \xi^0 & \chi^+ \\
		\chi^{--} & -\xi^- & \chi^ 0
	\end{pmatrix} ~,
\end{equation}
where the phase conventions of the charged fields are chosen to be $(\phi^+)^*=\phi^-$, $(\chi^+)^*=\chi^-$, $(\chi^{++})^*=\chi^{--}$, and $(\xi^+)^*=\xi^-$.

The most general scalar potential consistent with the SM gauge symmetry and the global $SU(2)_L\times SU(2)_R$ symmetry is given by
\begin{equation}\label{eq:potential}
\begin{aligned}
	V(\Phi,\Delta) &= \frac{m_1^2}{2}\Tr\left[\Phi^\dagger\Phi\right] + \frac{m_2^2}{2}\Tr\left[\Delta^\dagger\Delta\right] + \lambda_1\left(\Tr\left[\Phi^\dagger\Phi\right]\right)^2 + \lambda_2\left(\Tr\left[\Delta^\dagger\Delta\right]\right)^2 \\
	&\quad + \lambda_3\Tr\left[\left(\Delta^\dagger\Delta\right)^2\right] + \lambda_4\Tr\left[\Phi^\dagger\Phi\right]\Tr\left[\Delta^\dagger\Delta\right] + \lambda_5\Tr\left[\Phi^\dagger\frac{\sigma^a}{2}\Phi\frac{\sigma^b}{2}\right]\Tr\left[\Delta^\dagger T^a\Delta T^b\right] \\
	&\quad + \mu_1\Tr\left[\Phi^\dagger\frac{\sigma^a}{2}\Phi\frac{\sigma^b}{2}\right]\left(P^\dagger\Delta P\right)_{ab} + \mu_2\Tr\left[\Delta^\dagger T^a\Delta T^b\right]\left(P^\dagger\Delta P\right)_{ab} ~,
\end{aligned}
\end{equation}
where $\sigma^a$ and $T^a$ are the $2\times2$ and $3\times3$ representations of the $SU(2)$ generators, respectively,
and the matrix relating $\Delta$ to its Cartesian form is given by
\begin{equation}
	P = \frac{1}{\sqrt{2}}\begin{pmatrix}
		-1 & i & 0 \\
		0 & 0 & \sqrt{2} \\
		1 & i & 0
	\end{pmatrix} ~.
\end{equation}
The neutral fields are parametrized as $\phi^0=(v_\Phi+h_\phi+ia_\phi)/\sqrt{2}$, $\chi^0=(v_\Delta+h_\chi+ia_\chi)/\sqrt{2}$, and $\xi^0=v_\Delta+h_\xi$, where $v_{\Phi}$ and $v_\Delta$ denote their vacuum expectation values (VEVs).  Note that here $\langle \xi^0 \rangle = \sqrt2 \langle \chi^0 \rangle$ is required by the 
imposed global symmetry. The $SU(2)_L\times SU(2)_R \cong SU(2)_V\times SU(2)_A$ symmetry will be broken spontaneously by the VEVs down to the custodial $SU(2)_V$ symmetry, with $v=\sqrt{v_\Phi^2+8v_\Delta^2}\simeq246$~GeV. The two linearly independent minimum conditions
are given by
\begin{equation}\label{eq:tadpole}
\begin{aligned}
	m_1^2 &= -4\lambda_1v_\Phi^2 - 6\lambda_4v_\Delta^2-3\lambda_5v_\Delta^2-\frac{3}{2}\mu_1v_\Delta ~, \\
	m_2^2 &= -12\lambda_2v_\Delta^2 - 4\lambda_3v_\Delta^2 - 2\lambda_4v_\Phi^2 - \lambda_5v_\Phi^2 - \mu_1\frac{v_\Phi^2}{4v_\Delta} - 6\mu_2 v_\Delta ~.
\end{aligned}
\end{equation}

The scalar fields can be classified according to the isospin values of the
custodial $SU(2)_V$ symmetry into four multiplets:
\begin{equation}
\begin{split}
{\bf 5}:~ &
    H_5^{++} = \chi^{++} ~,~ 
    H_5^+ = \frac{1}{\sqrt{2}}\left(\chi^+-\xi^+\right) ~,~ 
    H_5 = \sqrt{\frac{1}{3}}h_\chi - \sqrt{\frac{2}{3}}h_\xi ~, \\
{\bf 3}:~ &
    H_3^+ = -\cos\beta~ \phi^+ + \frac{\sin\beta}{\sqrt{2}}(\chi^++\xi^+) ~,~ 
    H_3 = \cos\beta~ a_\phi + \sin\beta~ a_\chi ~, \\
{\bf 1}:~ &
    H_1 = \sin\alpha~ h_\phi + \frac{\cos\alpha}{\sqrt{3}}(\sqrt{2}h_\chi+h_\xi) ~, \\
{\bf 1}:~ &
    h = \cos\alpha~ h_\phi - \frac{\sin\alpha}{\sqrt{3}}(\sqrt{2}h_\chi+h_\xi) ~,
\end{split}
\end{equation}
where the mixing angle $\beta$ is defined through $\tan\beta=v_\phi/(2\sqrt{2}v_\Delta)$, and the other mixing angle $\alpha$ through
\begin{equation}
	\tan 2\alpha = \frac{2(M^2)_{12}}{(M^2)_{22}-(M^2)_{11}} ~,~~ 
	\alpha\in \left(-\frac{\pi}2, \frac{\pi}2 \right) ~,
\end{equation}
with
\begin{align}
\begin{split}
    (M^2)_{11} &= 8\lambda_1v^2\sin^2\beta ~, \\
    (M^2)_{22} &= (3\lambda_2+\lambda_3)v^2\cos^2\beta + M_1^2\sin^2\beta - \frac{1}{2}M_2^2 ~, \\
    (M^2)_{12} &= \sqrt{\frac{3}{2}}\sin\beta\cos\beta\left[(2\lambda_4+\lambda_5)v^2-M_1^2\right] ~,
\end{split}
\end{align}
and
\begin{equation}\label{eq:M1M2sq}
	M_1^2 = -\frac{v^2}{4v_\Delta}\mu_1 ,~ M_2^2 = -12 v_\Delta \mu_2 ~.
\end{equation}
The tree-level masses of the scalars are then given by
\begin{equation}
\begin{aligned}
        m_{H_5}^2 &= m_{H_5^\pm}^2 = m_{H_5^{\pm\pm}}^2 = M_1^2\frac{v_\Phi^2}{v^2} - \frac{3}{2}\lambda_5v_\Phi^2 + 8\lambda_3v_\Delta^2 + M_2^2 ~, \\
        m_{H_3}^2 &= m_{H_3^\pm}^2 = M_1^2-\frac{1}{2}\lambda_5v^2 ~, \\
        m_{H_1}^2 &= (M^2)_{11}\sin^2\alpha + (M^2)_{22}\cos^2\alpha + 2(M^2)_{12}\sin\alpha\cos\alpha ~, \\
        m_{h}^2 &= (M^2)_{11}\cos^2\alpha + (M^2)_{22}\sin^2\alpha - 2(M^2)_{12}\sin\alpha\cos\alpha ~.
\end{aligned}
\end{equation}
In the following, we will identify $h$ with the detected Higgs boson at about 125~GeV, and $H_1$ with the possible Higgs boson state at 95~GeV. Accordingly, we set $m_h=125$~GeV and $m_{H_1}=95$~GeV in the following.
In our numerical analysis, the parameter space preferred by the theoretical and experimental constraints that will be discussed below tends to have relatively small
$\alpha$ and $v_\Delta$ values (see Sec.~\ref{sec:numerical}). 
In the limit where $\alpha, v_\Delta \to 0$, known as the decoupling limit, the exotic scalar masses satisfy a decoupling mass relation, 
and their scale is driven to values far
above the electroweak scale~\cite{Chiang:2012cn}. Nevertheless, for the data points passing the applied constraints 
in our scan, this limit is not exactly realized. Instead, we find that
the mass relation mentioned above holds only approximately,
\begin{equation}\label{eq:decoup:masses}
    2m_{H_1}^2 \approx 3m_{H_3}^2 - m_{H_5}^2 ~,
\end{equation}
and that their values can be comparable to the electroweak scale.
Since the scale of all the exotic scalar masses is mainly set by the parameter $m_2^2$ (see Eq.~\eqref{eq:potential}), we expect from both this fact and from the approximate decoupling mass relation of Eq.~\eqref{eq:decoup:masses}
that the masses $m_{H_3}$ and $m_{H_5}$ should also be close to the electroweak scale, which points to the possibility of rich phenomenology with light BSM Higgs bosons. This has also been
studied in Ref.~\cite{Ahriche:2022aoj}.

We define the couplings of the SM weak gauge bosons (denoted by $V$) and charged fermions (denoted by $f$) to $h$ and $H_1$ in terms of $\kappa$ factors that modify the couplings as 
\begin{equation}
    g_{(h,H_1)(VV,ff)} = \kappa_{(h,H_1)(VV,ff)}\times g_{h(VV,ff)}^{\rm SM} ~,
    \label{eq:kappas}
\end{equation}
where in the lowest order
\begin{align}
\begin{split}
&
    \kappa_{hVV} = \sin\beta\cos\alpha - \sqrt{\frac{8}{3}}\cos\beta\sin\alpha ~,~~ 
    \kappa_{hff} = \frac{\cos\alpha}{\sin\beta} ~, \\
&
	\kappa_{H_1VV} = \sin\beta\sin\alpha + \sqrt{\frac{8} {3}}\cos\beta\cos\alpha ~,~~ 
	\kappa_{H_1ff} = \frac{\sin\alpha}{\sin\beta} ~.
\end{split}
\label{eq:kappaVVff}
\end{align}
We also define for the CP-odd state $H_3$ and 
the CP-even state $H_5$ their respective pseudoscalar and scalar couplings as
\begin{equation}
    g_{H_3ff} = {\rm sgn}(f)~ i \cot\beta~ g_{hff}^{\rm SM} ~,~ 
    g_{H_5VV} = -\frac{\cos\beta}{\sqrt{3}}g_{hVV}^{\rm SM} ~,
\end{equation}
where ${\rm sgn}(f=\text{up-type quarks}) = +1$ and ${\rm sgn}(f=\text{down-type quarks, charged leptons}) = -1$.  We note that the $\bf 3$ multiplet is gauge-phobic, whereas the $\bf 5$ multiplet is quark-phobic but can couple to leptons if lepton number-violating ($|\Delta L| = 2$) Yukawa couplings are introduced. Also, 
in the region of relatively small $\alpha$ and $v_\Delta$  
as described above, we find $\kappa_{h(VV,ff)}\sim 1$, 
as we will show in the numerical analysis in Sec.~\ref{sec:numerical},
while  $\kappa_{h(VV,ff)}=1$ holds in the exact decoupling limit.

In our study, we require the scalar potential to satisfy the following three sets of theoretical constraints at tree level:
\begin{description}
    \item[$\bullet~$Boundedness-from-below] The boundedness-from-below constraint can be satisfied as long as the quartic terms of the scalar potential remain positive for all possible field configurations. The sufficient and necessary conditions were first derived in Ref.~\cite{Hartling:2014zca}.
    
    \item[$\bullet~$Perturbative Unitarity] The perturbative unitarity constraint requires that the zeroth partial-wave mode of all $2\to2$ scattering channels should be smaller than $1/2$ at high energies. This was first studied and summarized in Ref.~\cite{Aoki:2007ah}.
    
    \item[$\bullet~$Unique Vacuum] The unique vacuum constraint requires that the custodially symmetric vacuum should be the unique global minimum in the scalar potential\footnote{We leave an investigation of
    the possibility that the custodially symmetric vacuum 
    could be only long-lived rather than being the global minimum
    within the context of the GM model 
    for future work 
    (see e.g.\ \citeres{Hollik:2018wrr,Ferreira:2019iqb} for such investigations in the MSSM and the N2HDM).}. This can be checked through numerically scanning different combinations of the triplet VEVs $\langle h_\chi\rangle$ and $\langle h_\xi\rangle$~\cite{Hartling:2014zca}.
    
    We remark that the assumption of misaligned triplet VEVs would break the custodial $SU(2)_V$ symmetry down to a $U(1)$ symmetry, resulting in undesired Goldstone bosons and tachyonic states~\cite{Chen:2022ocr}.  A more general scalar potential without the global $SU(2)_L\times SU(2)_R$ symmetry is required in order to consistently consider the scenario of having misaligned triplet VEVs~\cite{Blasi:2017xmc,Keeshan:2018ypw,Chen:2023ins}.
\end{description}
These constraints are applied to the potential parameters during the sampling process described in Sec.~\ref{sec:numerical}. For details of the implementation of these theoretical constraints, we refer to Ref.~\cite{Chen:2022zsh}.

\section{Numerical Analysis Setup}\label{sec:numerical}

\subsection{Experimental results for the 95-GeV excess}\label{subsec:95-GeV}

We quantify the compatibility of the model with the observed excesses at about 95~GeV using
\begin{equation}
    \chi^2_{\gamma\gamma,bb,\tau\tau} 
    = 
    \frac{(\mu_{\gamma\gamma,bb,\tau\tau}-\mu^{\rm exp}_{\gamma\gamma,bb,\tau\tau})^2}{(\Delta\mu_{\gamma\gamma,bb,\tau\tau}^{\rm exp})^2} ~,
\end{equation}
where the experimental central values and the uncertainties for the observed excesses in the $\gamma\gamma$, $bb$ and $\tau\tau$ channels are stated in Sec.~\ref{sec:intro}, and $\mu_{\gamma\gamma,bb,\tau\tau}$ are the theoretically predicted values for the signal strengths in the different channels.
Using the framework of coupling modifiers as defined in Eq.~\eqref{eq:kappas}, the theoretical predictions are given by
\begin{align}
    \mu_{\gamma\gamma}&\equiv\mu_{\gamma\gamma,{\rm LHC}}^{H_1} \equiv \kappa_{H_1ff}^2\times\frac{{\rm BR}(H_1\to \gamma\gamma)}{{\rm BR}(H_1\to \gamma\gamma)_{\rm SM}}  ~, \\
    \mu_{bb}&\equiv\mu_{bb,{\rm LEP}}^{H_1} \equiv \kappa_{H_1VV}^2\times\frac{{\rm BR}(H_1\to bb)}{{\rm BR}(H_1\to bb)_{\rm SM}}  ~, \\
    \mu_{\tau\tau}&\equiv\mu_{\tau\tau,{\rm LHC}}^{H_1} \equiv \kappa_{H_1ff}^2\times\frac{{\rm BR}(H_1\to \tau\tau)}{{\rm BR}(H_1\to \tau\tau)_{\rm SM}}  ~,
\end{align}
where we assume 100\% gluon-fusion production for $H_1$ at the LHC and denote the branching ratio for a
SM-like Higgs boson at 95~GeV to the final state $X$ as ${\rm BR}(H_1\to X)_{\rm SM}$. The lowest-order predictions for the coupling modifiers 
$\kappa_{H_1ff}$ and $\kappa_{H_1VV}$ are given in Eq.~\eqref{eq:kappaVVff}.

As mentioned in Sec.~\ref{sec:intro}, the di-tau excess 
observed by CMS at about $95 \gev$ is in some tension with other searches involving the di-tau final state.
On the other hand, the di-photon excess 
observed by CMS and ATLAS
appears to give rise to a more coherent picture. As a consequence,
we have chosen to analyze the experimental results
at three different levels: 1) $\gamma\gamma$, 2) $\gamma\gamma+bb$, and 3) $\gamma\gamma+bb+\tau\tau$. 
Thus, we only consider the di-photon excess in the first stage of our two-stage analysis framework (see below),
while the question of whether a simultaneous description of the excesses in the $bb$ and $\tau\tau$ channels is also possible is investigated 
in a separate step.

\subsection{Parameter scan and point selection}

The scalar potential of the GM model contains nine parameters: $m_{1,2}^2,\lambda_{1,2,3,4,5},\mu_{1,2}$ (see Eq.~\eqref{eq:potential}).  After fixing three of them via the constraints $m_h=125$~GeV, $m_{H_1}=95$~GeV, and $v\simeq246$~GeV, and trading one more of them with $v_\Delta$ through one of the minimum conditions (see Eq.~\eqref{eq:tadpole}), we are left with six independent degrees of freedom. For the numerical analysis, we choose the input parameters to be
\begin{equation*}
    \{v_\Delta ,~ \lambda_2 ,~ \lambda_3 ,~ \lambda_5 ,~ \mu_1 ,~ \mu_2\} ~,
\end{equation*}
and scan them uniformly within the range $v_\Delta\in[0,50]$~GeV, $\lambda_{2,3,5}\in[-4\pi,4\pi]$, $\mu_1\in[-500,0]$~GeV, and $\mu_2\in[-500,500]$~GeV.  We perform our numerical analysis in a two-stage framework. The first stage is to use the Bayesian-based Markov-Chain Monte Carlo simulation package \texttt{HEPfit}~\cite{DeBlas:2019ehy} to generate a collection of samples that are ``shaped'' by the applied theoretical and experimental constraints.  The second stage is to further constrain the allowed parameter space by applying the package \texttt{HiggsTools}~\cite{Bahl:2022igd} in order to ensure that the allowed parameter regions are in accordance with the measured properties of the detected Higgs boson at 125~GeV (sub-package \texttt{HiggsSignals}~\cite{Bechtle:2013xfa,Bechtle:2014ewa,Bechtle:2020uwn,Bahl:2022igd}, data set \texttt{v1.1}) and with the limits from searches for additional Higgs bosons at the LHC and at LEP (sub-package \texttt{HiggsBounds}~\cite{Bechtle:2008jh,Bechtle:2011sb,Bechtle:2013wla,Bechtle:2020pkv,Bahl:2022igd}, data set \texttt{v1.2}). 
The employed versions of \texttt{HiggsSignals} and \texttt{HiggsBounds} include 
essentially all relevant datasets
from the LHC Run~2.

In the first stage, after generating each sample from the previously defined scanning ranges, we calculate the quantities needed to test the theoretical constraints summarized in Sec.~\ref{sec:GM-model} as well as the total likelihood associated with all of the experimental measurements that we take into account at this stage, which include the 95-GeV di-photon excess, the 125-GeV Higgs rate measurements from LHC Run-1, and the $\mathrm{BR}(b\to s\gamma)$ measurement~\cite{ParticleDataGroup:2022pth} (following the methodology of \texttt{GMCALC}~\cite{Hartling:2014xma}).  All of the samples are then required to satisfy the theoretical constraints and to fall within the 95\% Bayesian confidence interval
\footnote{The $X$\% (Bayesian) confidence interval is defined as follows.  For a collection of samples that are individually assigned with a posterior probability, we sum up all these probabilities and sort the samples in the order of descending probability.  Then, starting from the first sample, the serial sub-collection of samples whose probability sum corresponds to $X$\% of the overall probability sum is considered to be within the ``$X$\% confidence interval''.}.
The reason that we only consider the 125-GeV Higgs rate measurements from LHC Run-1 at this stage is as follows. In both \texttt{HEPfit} and \texttt{HiggsTools}, the Run-1 measurements are all implemented in terms of signal strengths, i.e.\ as combinations of production and decay channels. Starting from Run-2, some of the measurements have been presented in terms of the STXS framework, 
which \texttt{HiggsTools} adopts
(where \texttt{HiggsSignals} ensures that no double counting of measurements occurs\footnote{
The code {\tt HiggsSignals} 
makes use of the LHC measurements for the 125-GeV Higgs boson in the detected final states.
These include the combined Run 1 rate~\cite{ATLAS:2016neq} measurements, the Run 2 measurements in the following final states,
$WW$~\cite{ATLAS:2018xbv,ATLAS:2019vrd,CMS:2020dvg}, 
$ZZ$~\cite{ATLAS:2020rej,CMS:2021ugl}, 
$b\bar{b}$~\cite{CMS:2018nsn,ATLAS:2020fcp,ATLAS:2020bhl,ATLAS:2021qou}, 
$c\bar{c}$~\cite{CMS:2019hve,ATLAS-CONF-2021-021}, 
$\tau^+\tau^-$~\cite{ATLAS:2018ynr,CMS-PAS-HIG-18-032,CMS:2021sdq,CMS:2022kdi}, 
$\mu^+\mu^-$~\cite{ATLAS:2020fzp,CMS:2020xwi}, 
$\gamma\gamma$~\cite{ATLAS-CONF-2020-026,CMS:2021kom,ATLAS:2022tnm},
as well as those measured in the $t\bar{t}H$ production channel~\cite{CMS:2018fdh,ATLAS-CONF-2019-045,CMS:2020mpn}.
Out of these the $\tau^+\tau^-$ and $\gamma\gamma$ final
states are incorporated via their respective STXS measurements, while for the others the measured signal strengths are implemented.
}), while \texttt{HEPfit} still uses the same framework as in Run-1. In order to apply the measurements consistently and since an exclusion of certain regions of the parameter space is only carried out in the second stage, we do not consider the Run-2 measurements at the first stage. This is also why we do not apply the limits from searches for additional Higgs bosons at this stage, which is to be considered in the next stage.

In the second stage, we further analyze the samples obtained in the previous stage, including the steps of: 
1) rejecting samples that violate the 95\% confidence level (CL) limits from experimental searches for additional Higgs bosons using 
\texttt{HiggsBounds}~\cite{Bechtle:2008jh,Bechtle:2011sb,Bechtle:2013wla,Bechtle:2020pkv,Bahl:2022igd}, and 2) calculating the total $\chi^2$ of the LHC rate measurements of the observed Higgs boson at 125~GeV, which we denote as $\chi^2_{125}({\rm GM})$, for the remaining samples using \texttt{HiggsSignals}~\cite{Bechtle:2013xfa,Bechtle:2014ewa,Bechtle:2020uwn,Bahl:2022igd}.  To evaluate the samples in more detail, we further calculate some additional quantities that are used in the later analysis.  Including $\chi^2_{125}({\rm GM})$, they are:
\begin{description}
    \item[$\bullet~\chi^2_{125}({\rm GM})$] The $\chi^2$ value associated with the rate measurements of the 125-GeV Higgs boson for the individual data points.
    \item[$\bullet~\chi^2_{X}({\rm GM})$] The $\chi^2$ value associated with the 95-GeV excesses in the $X=\gamma\gamma,\gamma\gamma+bb,\gamma\gamma+bb+\tau\tau$ channel(s) for the individual data points. For the details, please see the discussion in Secs.~\ref{sec:intro} and \ref{subsec:95-GeV}.
    \item[$\bullet~\chi^2_{125}({\rm SM})$] The $\chi^2$ value associated with the rate measurements of the 125-GeV Higgs boson for the case of the SM prediction.  We find $\chi^2_{125}({\rm SM}) \approx 152.5$.
    \item[$\bullet~\chi^2_{\rm X}({\rm SM})$] The $\chi^2$ value associated with the 95-GeV excesses in the $X=\gamma\gamma,\gamma\gamma+bb,\gamma\gamma+bb+\tau\tau$ channel(s) for the case of the SM.  
    We find $\chi^2_{\gamma\gamma}({\rm SM}) \approx 9.0$, $\chi^2_{\gamma\gamma+bb}({\rm SM}) \approx 13.2$, and $\chi^2_{\gamma\gamma+bb+\tau\tau}({\rm SM}) \approx 19.0$.
    \item[$\bullet~\Delta\chi^2_{125}$] The difference in the $\chi^2$ values associated with the rate measurements of the 125-GeV Higgs boson between the GM model and the SM, given by $\Delta\chi^2_{125} \equiv \chi^2_{125}({\rm GM})-\chi^2_{125}({\rm SM})$.
    \item[$\bullet~\Delta\chi^2_{X}$] $\Delta\chi^2_{X} \equiv 
    \Delta\chi^2_{125}
    +\chi^2_X({\rm GM})-\chi^2_{X}({\rm SM})$.
\end{description}
For each choice of $X$, we first select the subset of samples that satisfy $\Delta\chi^2_X<0$. For such samples, the combined $\chi^2$ arising from the excess at 95~GeV and the measurements of the Higgs boson at 125~GeV is lower than for the SM.  We then further pick from this subset the samples with $\Delta\chi^2_{125}<6.18$.  Such samples fall within the 95.4\% (or 2-sigma) CL of the SM prediction for the 125-GeV Higgs measurements for a two-dimensional parameter distribution. Finally, we identify the best-fit points from the latter subsets.


\section{Numerical analysis of the 95~GeV Higgs boson}\label{sec:95-results}

\begin{figure}[ht!]
    \centering
    \includegraphics[width=0.48\textwidth]{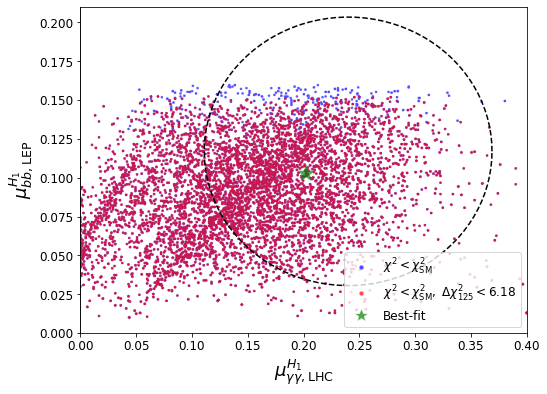}
    \includegraphics[width=0.48\textwidth]{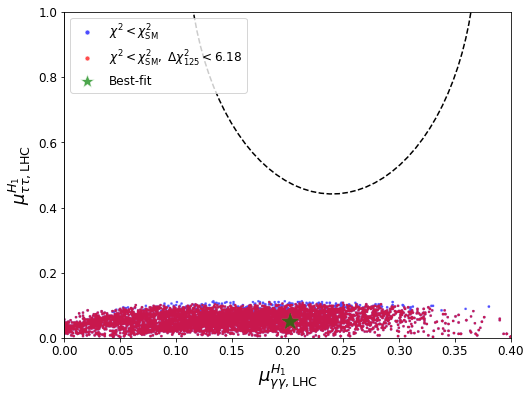}
    \\ \vspace{-0.3cm} \hspace{0.6cm} (a) \hspace{7.2cm} (b) \\[1em]
    \includegraphics[width=0.48\textwidth]{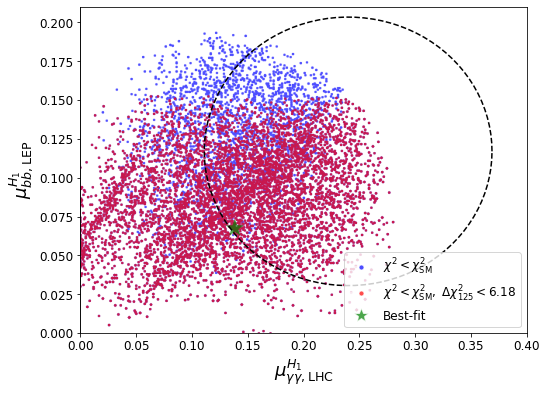}
    \includegraphics[width=0.48\textwidth]{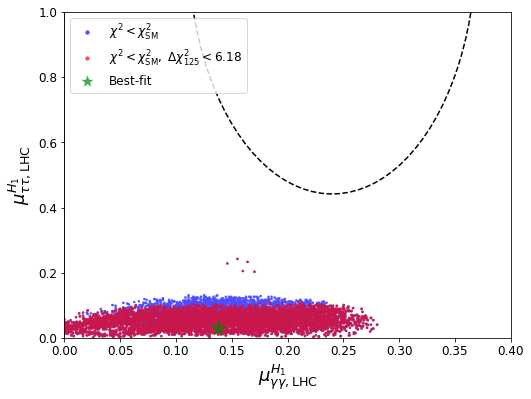}
    \\ \vspace{-0.3cm} \hspace{0.6cm} (c) \hspace{7.2cm} (d) \\[1em]
    \includegraphics[width=0.48\textwidth]{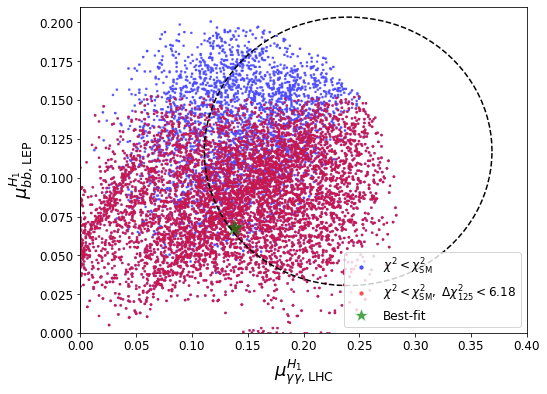}
    \includegraphics[width=0.48\textwidth]{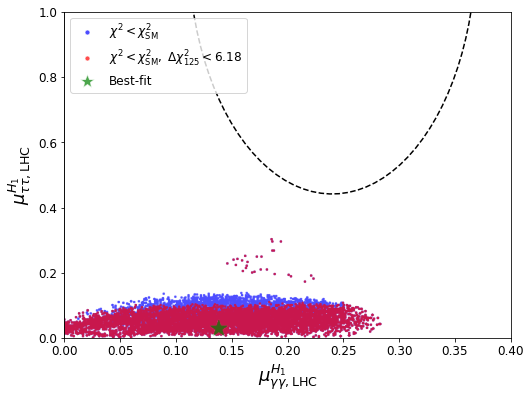}
    \\ \vspace{-0.3cm} \hspace{0.6cm} (e) \hspace{7.2cm} (f) \\[1em]
    \caption{\label{fig:mu_H1} Sample distributions in the $\mu_{H_1}^{\gamma\gamma}$--$\mu_{H_1}^{bb}$ plane for (a) $X=\gamma\gamma$ (case (i)), (c) $X=\gamma\gamma+bb$ (case (ii)), and (e) $X=\gamma\gamma+bb+\tau\tau$ (case (iii)), as well as in the $\mu_{H_1}^{\gamma\gamma}$--$\mu_{H_1}^{\tau\tau}$ plane for (b) case (i), (d) case (ii), and (f) case (iii). The elliptic contours denote the 1-$\sigma$ bounds of the corresponding 95-GeV excess measurements.}
\end{figure}

We first present in FIG.~\ref{fig:mu_H1} the sample distributions in the $\mu_{H_1}^{\gamma\gamma}$--$\mu_{H_1}^{bb}$ (left column) and 
$\mu_{H_1}^{\gamma\gamma}$--$\mu_{H_1}^{\tau\tau}$ planes (right column) for three cases: (i) $X=\gamma\gamma$ (upper row), (ii) $\gamma\gamma+bb$ (middle row) and (iii) $\gamma\gamma+bb+\tau\tau$ (lower row).
In all plots, we show the (black dashed) 1-sigma ellipses of the respective 95-GeV excesses.
The red points fulfill $\Delta\chi^2_X<0$ and $\Delta\chi^2_{125} < 6.18$. They are a subset of the blue points that
fulfill $\Delta\chi^2_X<0$.  The best-fit sample is marked by a green star.
One can see that in all three cases, the GM model cannot accommodate the $\tau\tau$ excess.  However, a large set of samples can be found within the 1-sigma contour of the $\mu_{H_1}^{\gamma\gamma}$--$\mu_{H_1}^{bb}$ plane.  It can also be seen that the $bb$ excess constraint restricts the data to about the left half of the $\gamma\gamma$ interval while still covering the central value.
 
We also point out that for the first case, the best-fit point is located well inside the 1-$\sigma$ contour, 
while for cases (ii) and (iii) it is located right at the boundary and, in particular, at lower $\mu_{\gamma\gamma}$ values. In view of the fact that the experimental situation regarding the $\tau\tau$ excess is somewhat unclear (see the discussion in Secs.~\ref{sec:intro} and \ref{subsec:95-GeV}), we will present only the results based upon the samples of case (i) in our further study.

\begin{table}[ht!]
    \centering
    \begin{tabular}{|>{\centering\arraybackslash}p{4.0cm}||>{\centering\arraybackslash}p{4.0cm}|>{\centering\arraybackslash}p{6.0cm}|}
    \toprule
    \multicolumn{3}{|c|}{Best-fit Point Properties} \\
    \toprule
    \multicolumn{3}{|c|}{$\alpha=0.224$,~$v_\Delta=5.24$~GeV,~$m_{H_3}=105$~GeV,~$m_{H_5}=121$~GeV} \\
    \colrule
    \multicolumn{3}{|c|}{$\kappa_{hVV}=0.952$,~$\kappa_{hff}=0.977$,~$\kappa_{H_1VV}=0.317$,~$\kappa_{H_1ff}=0.222$} \\
    \colrule
    \multirow{3}{*}{$\mu^{H_1}_{XX}$} & $\gamma\gamma$, LHC & 0.202 \\\cline{2-3}
    & $bb$, LEP & 0.103 \\\cline{2-3}
    & $\tau\tau$, LHC & 0.050 \\
    \colrule
    \multirow{9}{*}{${\rm BR}(h\to XX)$} & $bb$ & 0.581 \\\cline{2-3}
    & $cc$ & 0.029 \\\cline{2-3}
    & $\tau\tau$ & 0.062 \\\cline{2-3}
    & $gg$ & 0.094 \\\cline{2-3}
    & $\gamma\gamma$ & 0.002 \\\cline{2-3}
    & $Z\gamma$ & 0.002 \\\cline{2-3}
    & $ZZ$ & 0.025 \\\cline{2-3}
    & $WW$ & 0.204 \\
    \colrule
    \multirow{7}{*}{${\rm BR}(H_1\to XX)$} & $bb$ & 0.823 \\\cline{2-3}
    & $cc$ & 0.041 \\\cline{2-3}
    & $\tau\tau$ & 0.084 \\\cline{2-3}
    & $gg$ & 0.034 \\\cline{2-3}
    & $\gamma\gamma$ & 0.006 \\\cline{2-3}
    & $ZZ$ & 0.001 \\\cline{2-3}
    & $WW$ & 0.010 \\
    \colrule
    \multirow{4}{*}{${\rm BR}(H_3\to XX)$} & $bb$ & 0.781 \\\cline{2-3}
    & $cc$ & 0.039 \\\cline{2-3}
    & $\tau\tau$ & 0.081 \\\cline{2-3}
    & $gg$ & 0.098 \\\cline{2-3}
    \colrule
    \multirow{4}{*}{${\rm BR}(H_5\to XX)$} & $WW$ & 0.546 \\\cline{2-3}
    & $ZZ$ & 0.208 \\\cline{2-3}
    & $\gamma\gamma$ & 0.225 \\\cline{2-3}
    & $Z\gamma$ & 0.021 \\\cline{2-3}
    \colrule
    ${\rm BR}(H_3^\pm\to \tau\nu_\tau)$ & $1.0$ & \multirow{3}{*}{\shortstack{\hfill\\Off-shell scalar final states \\are neglected.}} \\\cline{1-2}
    ${\rm BR}(H_5^\pm\to WZ)$ & $1.0$ & \\\cline{1-2}
    ${\rm BR}(H_5^{\pm\pm}\to WW)$ & $1.0$ & \\
    \botrule
    \end{tabular}
    \vspace{1em}
    \caption{Summary of the physical properties of the best-fit point in the $X = \gamma\gamma$ selection.}
    \label{tab:best-fit}
    \vspace{2em}
\end{table}

A summary of the physical properties of the best-fit point for $X=\gamma\gamma$ is given in TABLE.~\ref{tab:best-fit}. 
While this is only an example of one benchmark point, it gives an idea about phenomenologically preferred parameters and decay channels around that region of the parameter space.  One can observe that $\alpha=0.224$ and $v_\Delta=5.24$~GeV (thus $\cos\beta\approx 0.06$), while the values $m_{H_3}=105$~GeV and $m_{H_5}=121$~GeV show that the mass spectrum of the scalar states lies far below the TeV scale and that the approximate decoupling mass relation $3m_{H_3}^2-m_{H_5}^2\approx 2 \, (95~{\rm GeV})^2$ (see Eq.~\eqref{eq:decoup:masses}) is satisfied. These characteristics demonstrate that the preferred parameter space tends to yield $\kappa_{h(VV,ff)} \approx 1$, which for the best-fit point have the values $\kappa_{hVV}=0.952,~\kappa_{hff}=0.977$.  Given the masses of the exotic scalars, the decay of $H_3$ is dominated by the $bb$ channel and that of $H_3^\pm$ by the $\tau\nu_\tau$ channel.  Though the $H_5$ boson primarily decays into the $WW$ final state, the branching ratios of the $\gamma\gamma$ and $ZZ$ channels are also of the same order. As we will show later, the preferred mass range of $H_5$ in our samples actually covers 
the three regimes of $m_{H_5}\lesssim 2m_W$, $2m_W\lesssim m_{H_5}\lesssim 2m_Z$, and $2m_Z\lesssim m_{H_5}$, thus resulting in a diverse combination of $\gamma\gamma,WW,ZZ$ channel preferences. However, the production cross sections for $H_5$ 
obtained from our samples are all $\lesssim\mathcal{O}(1~{\rm fb})$,
which limits the prospects for probing this state at the LHC in the near future. 
Finally, though we do not consider their off-shell decays to other scalar states, $H_5^\pm$ and $H_5^{\pm\pm}$ primarily decay into $WZ$ and same-sign $WW$ final states, respectively, which are the striking signatures of these GM scalar states.

\begin{figure}[ht!]
    \centering
    \includegraphics[width=0.48\textwidth]{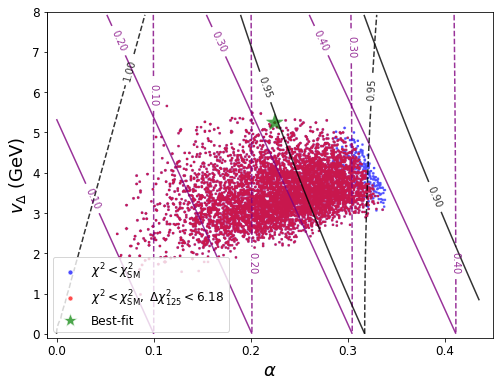}
    \includegraphics[width=0.40\textwidth]{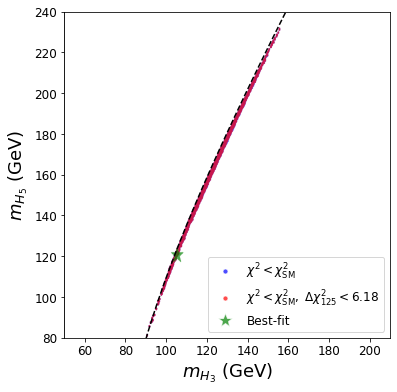}
    \\ \vspace{-0.3cm} \hspace{0.6cm} (a) \hspace{7.2cm} (b) \\
    \caption{\label{fig:GM:phys} Sample distributions in the (a) $\alpha$--$v_\Delta$ and the (b) $m_{H_3}$--$m_{H_5}$ planes for case (i). In plot (a), the solid (dashed) lines denote the $\kappa_{XVV}$ ($\kappa_{Xff}$) contours for $X=h$ and $X = H_1$ in black and purple, respectively.  In plot (b), the dashed curve denotes the contour of the coupling mass relation given by Eq.~\eqref{eq:decoup:masses} in the exact limit.
    }
\end{figure}

In FIG.~\ref{fig:GM:phys} we present the sample distributions in the $\alpha$--$v_\Delta$ and $m_{H_3}$--$m_{H_5}$ planes. In FIG.~\ref{fig:GM:phys}(a), we also plot the contours of $\kappa_{(h,H_1)(VV,ff)}$: the solid (dashed) lines denote the $\kappa_{XVV}$ ($\kappa_{Xff}$) contours for $X=h$ and $X = H_1$ in black and purple, respectively.  In FIG.~\ref{fig:GM:phys}(b), we also indicate the contour of the approximate decoupling mass relation given by Eq.~\eqref{eq:decoup:masses} using a black dashed curve. 
In contrast to the case of $m_{H_1}>m_h$, where one finds $\alpha<0$ (see, e.g., Ref.~\cite{Chen:2022zsh}), in our analysis with 
$95\gev = m_{H_1} < m_h = 125\gev$, all the data points are found to have $\alpha>0$. Furthermore, one can see that 
$\alpha \lesssim 0.35$ and $v_\Delta \lesssim 6$~GeV, and thus the magnitudes of the $\kappa$ values manifest the following features of the small $\alpha$ and $v_\Delta$ limit: the $\kappa_{h(VV,ff)}$ values are close to the SM predictions while $\kappa_{H_1(VV,ff)} \lesssim 0.4$. 
These are further confirmed by plot (b), where all points are found to be very close to the contour indicating the decoupling mass relation given by Eq.~\eqref{eq:decoup:masses} in the exact limit. The feature that the preferred parameter region  
yields $\kappa_{h(VV,ff)}$ values that are close to the SM predictions
can be understood from the fact that all the current measurements of the properties of the Higgs boson at
125~GeV are very consistent with the SM predictions, thus leaving little space for the model parameters to deviate from this limit. Nevertheless, in the GM model the additional triplet gauge couplings provide more flexibility (parametrized by $\beta$) 
in comparison to the case of just a
singlet scalar mixing (parametrized by $\alpha$) to account for both the 125- and 95-GeV Higgs measurements, while also allowing rich phenomenology in the other sectors of the model. Moreover, the smallness of $m_{H_3}$ and $m_{H_5}$ ($\lesssim160$ and $230$~GeV, respectively), which is expected from the approximate decoupling mass relation of Eq.~\eqref{eq:decoup:masses} and the general scale set by $m_2^2$ as discussed in Sec.~\ref{sec:GM-model}, suggests that in the GM model the confirmation of the observed excess at about 95~GeV would give rise to exciting prospects regarding possible discoveries of further sub-TeV Higgs bosons at the (HL-)LHC (see the discussion below). As alluded to before, the preferred $m_{H_5}$ range of $[90,230]$~GeV covers both the $2m_W$ and $2m_Z$ thresholds and thus gives rise to various $H_5$ decay patterns, but the smallness of its production cross section renders it 
difficult
to be probed in the near future experiments.

\begin{figure}[ht!]
    \centering
    \includegraphics[width=0.48\textwidth]{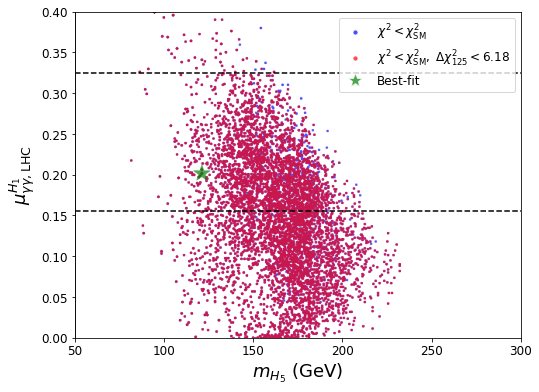}
    \includegraphics[width=0.48\textwidth]{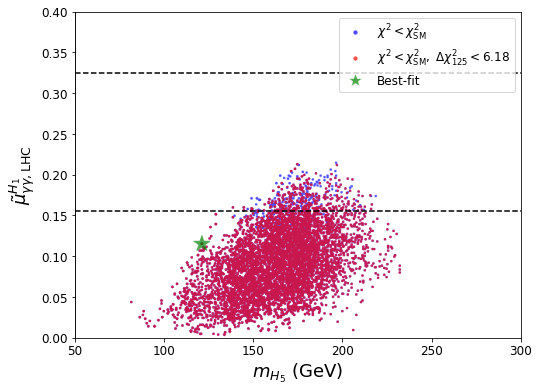}
    \\ \vspace{-0.3cm} \hspace{0.6cm} (a) \hspace{7.2cm} (b) \\
    \caption{\label{fig:mu:mH5} Sample distributions in the (a) $m_{H_5}$--$\mu^{H_1}_{\gamma\gamma}$ and (b) $m_{H_5}$--$\tilde{\mu}_{\gamma\gamma}^{H_1}$ planes for case (i).}
\end{figure}

In the final step of our analysis of the present experimental situation, we investigate the origin of the relatively large BR($H_1 \to \gamma\gamma$), as required to fit the CMS and ATLAS excesses.  Here the $H_5^{\pm\pm}$-loop diagram contribution to the effective $H_1\gamma\gamma$ coupling plays an important role.  To study this, we define $\tilde{\mu}^{H_1}_{\gamma\gamma}$ to be the 95-GeV di-photon signal strength predicted by the GM model ``without the $H_5^{\pm\pm}$-loop contribution''.  The comparison of the results with and without the loop contribution involving $H_5^{\pm\pm}$ is shown in FIG.~\ref{fig:mu:mH5}, where we plot the sample distributions in both the $m_{H_5}$--$\mu^{H_1}_{\gamma\gamma}$ and $m_{H_5}$--$\tilde{\mu}_{\gamma\gamma}^{H_1}$ planes. One can clearly see that removing the $H_5^{\pm\pm}$-loop diagram significantly lowers the 95-GeV di-photon signal strength predictions.  This applies especially to points with relatively low $m_{H_5}$, as the doubly-charged Higgs boson loop contribution is larger for lighter $H_5^{\pm\pm}$. This feature of the GM model (and some other triplet extensions such as the Type-II seesaw model~\cite{Magg:1980ut,Cheng:1980qt,Lazarides:1980nt,Mohapatra:1980yp}) thus motivates the search for a relatively light doubly-charged scalar boson in future experiments. One such search was performed by ATLAS within the context of the GM model~\cite{ATLAS-CONF-2023-023}, which reported a $2.5\,\sigma$ excess of $H_5^{\pm\pm}$ at 450~GeV in the VBF production channel. While this is far above the preferred $m_{H_5}$ range of our present study, it motivates further dedicated searches for doubly-charged Higgs bosons also for smaller masses at the LHC in the future.

\section{Future prospects}\label{sec:future}

\begin{figure}[ht!]
    \centering
    \includegraphics[width=0.48\textwidth]{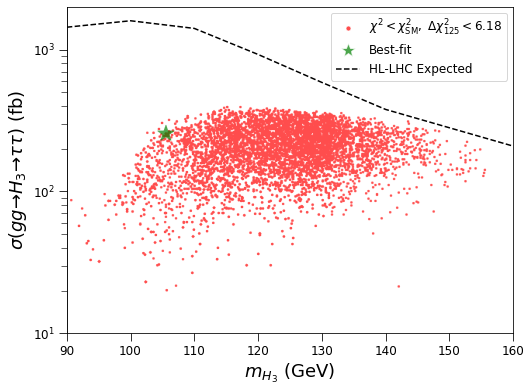}
    \includegraphics[width=0.48\textwidth]{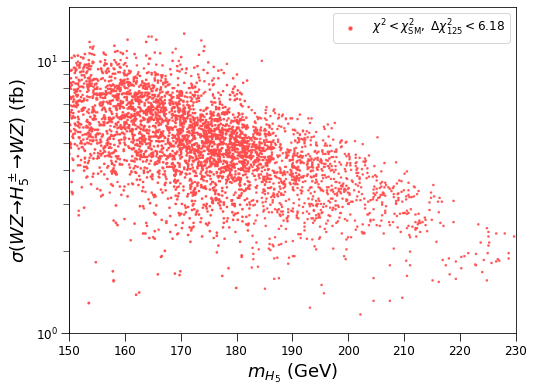}
    \\ \vspace{-0.3cm} \hspace{0.6cm} (a) \hspace{7.2cm} (b) \\
    \includegraphics[width=0.48\textwidth]{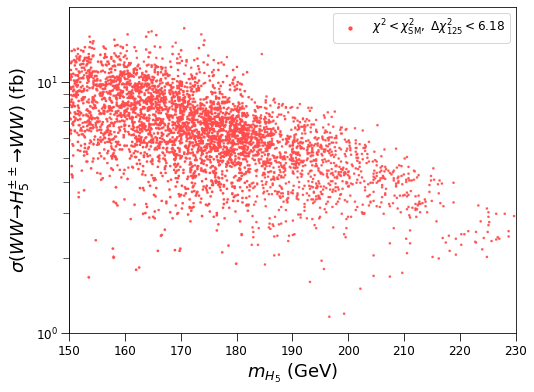}
    \\ \vspace{-0.3cm} \hspace{0.6cm} (c)  \\
    \caption{\label{fig:xsecs} Sample distributions in the (a) $m_{H_3}$--$\sigma(gg\to H_3\to\tau\tau)$, (b) $m_{H_5}$--$\sigma(WZ\to H_5^\pm\to WZ)$, and (c) $m_{H_5}$--$\sigma(WW\to H_5^{\pm\pm}\to WW)$ planes at the 14-TeV LHC for case (i). In (a) we further show the expected 95\% CL HL-LHC exclusion bound~\cite{Cepeda:2019klc}.}
\end{figure}

We finish our analysis by investigating the prospects for testing
the GM interpretation of the excesses at 95~GeV. We start with an analysis
of the most promising search channels at the HL-LHC in FIG.~\ref{fig:xsecs}. 
The upper left plot shows our sample with all points fulfilling $\chi^2 < \chi^2_{\rm SM}$ and 
$\Delta\chi^2_{125} < 6.18$ in the plane
$m_{H_3}$--$\sigma(gg\to H_3\to\tau\tau)$\footnote{The cross sections used in this section are all obtained from the predictions
reported by the LHC Higgs Cross Section Working Group~\cite{LHCHiggsCrossSectionWorkingGroup:2013rie,LHCHiggsCrossSectionWorkingGroup:2016ypw}, using the $\kappa$ parameters as specified in Sec.~\ref{sec:GM-model} as coupling modifiers.}, where the black dashed line indicates
the anticipated 95\%~CL HL-LHC search limit for the search of Higgs bosons decaying to $\tau^+\tau^-$~\cite{Cepeda:2019klc}.
It can be seen that according to the current projections for the achievable
sensitivity, it will not be possible to discover $H_3$ in the $\tau^+\tau^-$ decay channel. Only with a substantially improved sensitivity will this channel become accessible as a discovery mode at the HL-LHC.
The upper right and the lower plot show our sample points in the $m_{H_5}$--$\sigma(WZ\to H_5^\pm\to WZ)$ and
$m_{H_5}$--$\sigma(WW\to H_5^{\pm\pm}\to WW)$ planes, respectively. No evaluation of the HL-LHC reach is available for these search channels. However, with masses around and not too far below the $WZ$ and $WW$ mass shells and cross sections above 1~fb, it may be possible to cover 
part of the parameter space at the HL-LHC.

\begin{figure}[ht!]
    \centering
    \includegraphics[width=0.73\textwidth]{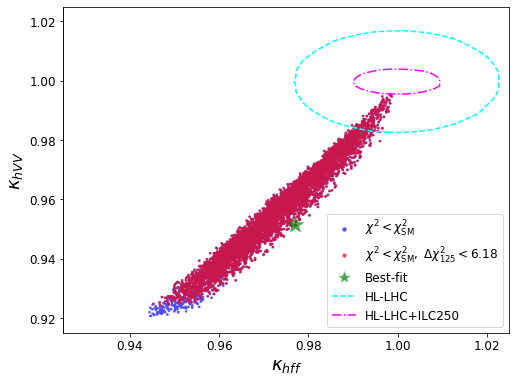}
    \caption{\label{fig:khff-khVV} Sample distribution in the $\kappa_{hff}$--$\kappa_{hVV}$ plane for case (i) and the prospective precision at the 
    $1\,\sigma$ level (indicated for the SM value)
    at the HL-LHC (cyan)~\cite{Cepeda:2019klc} and the HL-LHC+ILC250 (magenta)~\cite{Bambade:2019fyw}.}
\end{figure}

\begin{figure}[ht!]
    \centering
    \includegraphics[width=0.73\textwidth]{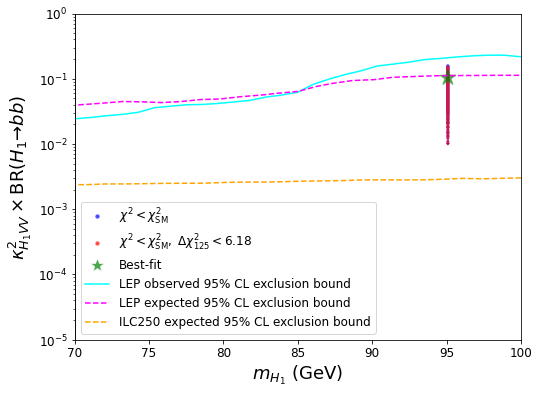}
    \caption{\label{fig:mH1-mu_H1_bb} Sample distribution in the $m_{H_1}$--$\kappa_{hVV}^2\times{\rm BR}(H_1\to bb)$ plane for case (i) and the 95\% CL LEP observed exclusion bound~\cite{LEPWorkingGroupforHiggsbosonsearches:2003ing} (cyan), the 95\% CL LEP expected exclusion bound~\cite{LEPWorkingGroupforHiggsbosonsearches:2003ing} (magenta), and a projection for the 95\% CL ILC250 expected exclusion bound~\cite{Drechsel:2018mgd} (orange).}
\end{figure}

\begin{figure}[ht!]
    \centering
    \includegraphics[width=0.7\textwidth]{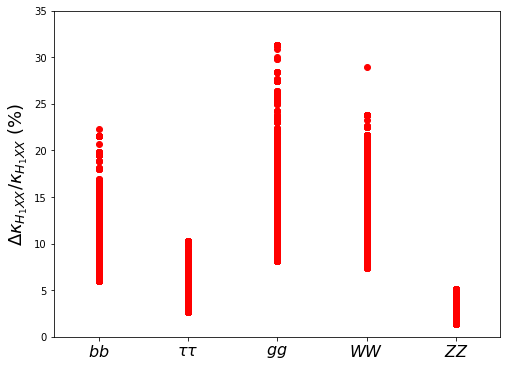}
    \caption{\label{fig:Delta-kH1} Sample distributions for case (i) of $\Delta\kappa_{H_1XX}/\kappa_{H_1XX},~XX=bb,\tau\tau,gg,WW,ZZ$.}
\end{figure}

Before moving on, we comment on two 
types of $H_5^{\pm\pm}$ searches that turn out to impose no or incomplete constraints on the $m_{H_5}$ range in our samples. The first 
type of searches are the ones for the process $H_5^{\pm\pm}\to\ell^\pm\ell^\pm$. In the GM model, this process is realized through the lepton number-violating coupling between the $\chi$ and lepton fields. However, Ref.~\cite{Chiang:2012cn} shows that only 
for $v_\Delta\lesssim10^{-4}$~GeV 
would this decay channel become comparable to the weak gauge boson decay channels, and thus it is irrelevant 
for the preferred parameter space of 
our study. One such search has been performed at LEP, targeting the pair production of $H_5^{\pm\pm}$~\cite{OPAL:2001luy}.
The second 
type of searches are the ones for $H_5^{\pm\pm}$ relying on the decay to two on-shell like-sign $W$ bosons, which have been performed at the LHC and
resulted in
a stringent bound of $m_{H_5}\geq 200$~GeV~\cite{CMS:2017fhs,ATLAS:2018ceg,ATLAS:2021jol,CMS:2021wlt}.
Therefore, it remains 
an open question
to which extent a doubly-charged Higgs boson
$H_5^{\pm\pm}$ with a mass below 200~GeV
can be probed
at the LHC. Recently, a new method was presented, targeting this doubly-charged Higgs-boson
mass scale~\cite{Ashanujjaman:2022ofg}.
The new search strategy focuses on $H^{\pm\pm}$ pair production in the highly
boosted regime, i.e., on $H^{++}H^{--}$ pairs with large $p_T$, with one of them decaying via a pair of same-sign $W$ bosons to a fat jet and the other decaying via another pair of same-sign $W$ bosons to two
adjacent same-sign leptons and two neutrinos, i.e., $p_T^{\rm miss}$.
While the requirement of large $p_T$ results in a significant reduction of the
number of signal events, the SM background is even further suppressed,
particularly via a discrimination of the fat jet from SM background jets.
Depending on the mass of the doubly-charged Higgs boson,
\citere{Ashanujjaman:2022ofg} claims to 
have sensitivity 
to discover doubly-charged Higgs bosons with less
than $\sim 160$~fb$^{-1}$ of LHC data, in the mass range between $100 \gev$ and
$200 \gev$. While no such search, employing all the currently available data, has been
performed so far, this indicates that
the GM model interpretation of the excesses around $\sim 95 \gev$ may be testable in the upcoming 
LHC Runs.

Finally, we analyze the potential of future $e^+e^-$ colliders to further probe the GM interpretation of the 
95~GeV excesses.
We first present in FIG.~\ref{fig:khff-khVV} the sample distribution in the $\kappa_{hff}$--$\kappa_{hVV}$ plane 
overlaid with the anticipated precisions for the coupling measurements at the HL-LHC (cyan dashed)~\cite{Cepeda:2019klc}
and combined with (hypothetical future) ILC250 measurements (magenta dot-dashed)~\cite{Bambade:2019fyw}. 
The ellipses in the plot are centered around the SM prediction of $\kappa_{hff} = \kappa_{hVV} = 1$.
The deviations predicted in the $h$~couplings w.r.t.\ the SM predictions can be very small (see the discussion above). Consequently,
a sizable fraction of the sample points are within the $1\,\sigma$ HL-LHC ellipse, and only the largest deviations would yield 
a $\ge 3\,\sigma$ distinction of the GM and the SM. 
The situation is substantially improved for the case where the prospective
ILC250 measurements are included.
However, even including the $e^+e^-$ coupling measurement a relevant part of the predicted points lies within 
$2\,\sigma$ of the SM prediction. 
In the final step, we analyze the capabilities of the ILC to produce the new Higgs boson at $\sim 95$~GeV, i.e.~$H_1$, 
and to measure its couplings. In FIG.~\ref{fig:mH1-mu_H1_bb} we show the plane 
$m_{H_1}$--$\kappa_{H_1} \times {\rm BR}(H_1 \to b\bar b)$. The cyan (magenta) line indicates the observed (expected) 
exclusion obtained at LEP~\cite{LEPWorkingGroupforHiggsbosonsearches:2003ing}, where the $\sim 2\,\sigma$ excess around 
$95-98$~GeV can be seen. The dashed orange line indicates the improvements that can be expected at the ILC250 with 
an integrated luminosity of 2~ab$^{-1}$ according to the projection of Ref.~\cite{Drechsel:2018mgd} (see also Ref.~\cite{Wang:2018fcw}). 
The blue points (which are a superset of the red points) show our selected sample, i.e.\ with $\chi^2 < \chi^2_{\rm SM}$, where the red points furthermore
fulfill $\Delta\chi^2_{125} < 6.18$. It can clearly be observed that all 
parameter points within the preferred region are well within the projected 
ILC250 sensitivity. Consequently, it is expected within the GM model that the new Higgs boson at $\sim 95$~GeV can be produced abundantly at the ILC250 (or other $e^+e^-$ colliders operating at $\sqrt{s} = 250$~GeV). 
In FIG.~\ref{fig:Delta-kH1} we analyze the prospects of $H_1$ coupling measurements at the ILC250. The evaluation of the anticipated precision of the coupling measurements is based on Ref.~\cite{Heinemeyer:2021msz}. 
The evaluation has been performed for the (effective) couplings of the $H_1$ to $b\bar b$, $\tau^+\tau^-$, $gg$, 
$WW$ and $ZZ$. While the first four rely on the decay of the Higgs boson to the respective final state, the $H_1ZZ$ 
coupling is obtained from the production of the $H_1$ as radiated from a $Z$~boson. The latter channel yields the highest
precision between $1-5\%$. A high accuracy is also expected for the coupling to $\tau$-leptons, ranging from $2-10\%$. 
The other three couplings are expected to be determined with an accuracy between $\sim 6\%$ and $\sim 30\%$. Coupling measurements at this level of precision will help to distinguish the GM interpretation of the 95~GeV excesses from other model interpretations, see, e.g., Refs.~\cite{Heinemeyer:2021msz,Biekotter:2022jyr}, where prospective coupling precisions for the N2HDM, 2HDMS and S2HDM have been evaluated.

\section{Conclusions}
\label{sec:conclusions}

If confirmed by further data, the excesses reported by CMS and ATLAS in the searches for low-mass Higgs bosons in the di-photon decay channel at a mass value of about $95\gev$ could constitute a direct manifestation of an extended Higgs sector via the production of a new Higgs boson.
In many previous model interpretations of the observed excesses in terms of a state $\phi$, extensions of the 2HDM were employed. 
This was mainly due to the possibility of a suppressed
$\phi b\bar b$ coupling, thereby enhancing BR($\phi \to \gamma\gamma$) 
in such a way that the CMS and ATLAS excesses in this channel can be properly described. In the present paper, we have investigated a different  
possibility for enhancing the di-photon decay rate.  Additional charged 
particles in the loop-mediated decay of a $\sim 95 \gev$ Higgs boson to two photons can yield a positive contribution 
to the decay rate and in this way result in a sufficiently strong rate. 
However, it was demonstrated in \citere{Biekotter:2019kde} that a second Higgs doublet, providing an additional 
singly-charged scalar, is not sufficient to yield a relevant effect on BR($\phi \to \gamma\gamma$). 

The situation is different in BSM models that extend the Higgs sector with triplets.  Such models predict the existence of doubly-charged Higgs bosons, which can substantially contribute to the di-photon decay rate of a new light, neutral Higgs boson that is also allowed to exist in the model.
The GM model is of particular interest in this context, since despite containing Higgs triplets, 
it preserves the electroweak $\rho$-parameter to be~1 at the tree level. 
We analyzed the di-photon excess within the GM model in conjunction with an excess in the
$b \bar b$ final state observed at LEP and an excess observed by CMS in the di-tau final state, which were found at comparable masses with local significances of about $2\sigma$ and $3\sigma$, respectively. 
We demonstrated that within the GM model, a $\sim 95 \gev$ Higgs boson with a di-photon decay rate as observed by CMS and ATLAS 
can be well described. Simultaneously,
the GM model can also accommodate
the $b \bar b$ excess at LEP, but not the di-tau excess.
In this context, it is important to note that the signal strength observed by CMS in the $gg \to \phi \to \tau^+\tau^-$
search is in some tension (depending on the model realization) with experimental bounds from
recent searches performed by CMS for the production of a Higgs boson
in association with a top-quark pair or in association with a $Z$~boson, with subsequent decay into tau pairs, as well as with the searches performed at LEP for the process $e^+e^-\to Z\phi(\phi\to\tau^+\tau^-)$.

We have demonstrated in our analysis that the loop contribution of the doubly-charged Higgs boson indeed gives rise to an important upward shift in the 
di-photon rate of the Higgs boson at $\sim 95 \gev$ which is instrumental for bringing the predicted rate in agreement with the observed excesses. In the preferred parameter region, the doubly-charged Higgs boson is predicted to be rather light, with a mass in the range between $100 \gev$ and $200 \gev$. While the LEP searches excluded doubly-charged Higgs bosons below $\sim 100 \gev$ using the di-tau final state, which is irrelevant to the preferred parameter space of this study, the searches that were conducted at the LHC so far only cover the mass region above $\sim 200 \gev$. 

Since also the other Higgs bosons of the GM model are predicted to be light in the considered scenario,
a realization of the observed excesses at
$\sim 95 \gev$ within the GM model would give rise to the exciting possibility that experimental confirmation of the excesses at $\sim 95 \gev$ could be accompanied by discoveries of further states of the extended Higgs sector. We have studied the prospects for probing the GM interpretation with results from future Runs of the LHC and from a future $e^+e^-$ collider.
While dedicated searches for low-mass doubly-charged Higgs bosons at the LHC could probe the preferred mass region, we find that
the predicted rates for the production of
Higgs bosons decaying to tau-pairs remain below the anticipated reach of the HL-LHC.
A future $e^+e^-$ collider would have good prospects for probing the GM interpretation of the observed excesses at $\sim 95 \gev$ via precision measurements of the couplings of the detected Higgs boson at about $125\gev$, via the direct search for the state at $\sim 95 \gev$ and via searches for the pair production of the doubly-charged Higgs boson of the GM model.

\section*{Acknowledegments}

We thank
S. Ashanujjaman
and
T.~Biek\"otter 
for helpful discussions.
C.W.C.\ is supported in part by the National Science and Technology Council of Taiwan under Grant No.~NSTC-111-2112-M-002-018-MY3.
The work of S.H.\ has received financial support from the
grant PID2019-110058GB-C21 funded by
MCIN/AEI/10.13039/501100011033 and by ``ERDF A way of making Europe", 
and in part by the grant IFT Centro de Excelencia Severo Ochoa CEX2020-001007-S
funded by MCIN/AEI/10.13039/501100011033. 
S.H.\ also acknowledges support from Grant PID2022-142545NB-C21 funded by
MCIN/AEI/10.13039/501100011033/ FEDER, UE.
G.W.\ acknowledges support by the
Deutsche Forschungsgemeinschaft (DFG, German Research Foundation) under Germany’s
Excellence Strategy~–~EXC 2121 “Quantum Universe”~–~390833306. This work has been partially funded by the Deutsche Forschungsgemeinschaft (DFG, German Research Foundation)~-~491245950.

\bibliography{reference}

\end{document}